\documentstyle[preprint,aps]{revtex}
\begin{document}
\draft
\title{A New Test of the Einstein Equivalence Principle\\
        and the Isotropy of Space}
\author{Mark P. Haugan and Thierry F. Kauffmann }
\address{Department of Physics, Purdue University 1396, West Lafayette
	IN 47907}
\date{}
\maketitle

\begin{abstract}
Recent research has established that nonsymmetric gravitation theories like
Moffat's NGT predict that a gravitational field singles out an orthogonal
pair of polarization states of light that propagate with different phase
velocities. We show that a much wider class of nonmetric theories encompassed
by the $\chi g$ formalism predict such violations of the Einstein equivalence
principle. This gravity-induced birefringence of space implies that
propagation through a gravitational field can alter the polarization of
light. We use data from polarization measurements of extragalactic sources
to constrain birefringence induced by the field of the Galaxy. Our new
constraint is $10^8$ times sharper than previous ones.
\end{abstract}


\section{Introduction}
Gabriel\cite{Gabriel1} {\it et al}.\ recently established that members of a
certain class of nonmetric gravitation theories predict that space is
anisotropic and birefringent.  These nonsymmetric theories, for which
Moffat's NGT\cite{Moffatt} is the prototype, have been studied extensively
as potentially viable alternatives to general relativity, but they predict
that a gravitational field singles out an orthogonal pair of linear
polarization states of light which propagate through the field with different
phase velocities.  The difference in phase velocities could, in principle,
be measured in local test experiments\cite{Gabriel1} and, so, violates
the Einstein equivalence principle.  In practice, the sharpest constraints
on the magnitude of such gravity-induced birefringence will be inferred
from limits on the cumulative effect that propagation through a
gravitational field has on the polarization of light.  Gabriel\cite{Gabriel3}
{\it et al}.\ consider the effect that the Sun's field could induce and
use measurements of the polarization of solar spectral lines to impose sharp
new constraints on nonsymmetric gravitation theories.  Krisher\cite{Krisher}
considers the effect the Galaxy's field could induce and uses pulsar
polarization observations to impose a complementary constraint on NGT.

In this paper we show that the class of nonmetric gravitation theories
which predict spatial anisotropy or birefringence is far more extensive
than the class of nonsymmetric theories.  Indeed, one must take pains
to define a nonmetric coupling between the gravitational and electromagnetic
fields that does not induce anisotropy and birefringence.  Consequently,
searching for effects of propagation through a gravitational field on the
polarization of light provides tests of the Einstein equivalence principle
that have a power, generality and significance comparable to those of
more familiar atomic anisotropy (Hughes-Drever) tests\cite{Hughes-Drever}
 and E\"otv\"os and gravitational redshift tests\cite{Dicke}. Recent
theoretical developments provide a strong motivation for precise new tests
of the Einstein equivalence principle.  For example, interest in scalar-tensor
and multi-tensor gravitation theories has been rekindled by the proliferation
of scalar and tensor fields in effective field theories derived from string
physics and several threads of quantum gravity research suggest that
connection rather than metric may be fundamental.

The view that searching for evidence of gravity-induced birefringence
tests the Einstein equivalence principle is not unprecedented.  Ten years ago
Ni\cite{Ni8} remarked that nonmetric gravitation theories encompassed by his
$\chi g$ formalism could predict birefringence of the type we study.  He also
noted that pulsar polarization observations could constrain this possibility.
Perhaps the significance of these comments was overlooked at the time because
no gravitation theories predicting such birefringence were known and because
the effective time resolution of pulsar polarization observations is limited.
In any event, their significance is now clear.  Theories like NGT do predict
gravity-induced birefringence and we use galaxy polarization observations to
constrain this possibility far more stringently than Ni or Krisher were able
to using pulsar polarization observations.  Indeed, the linear and circular
polarization data for galaxies that we analyze have an effective time
resolution that is more than $10^8$ times that of Krisher's pulsar
polarization data.

\section{ Electrodynamics in a Background Gravitational Field}

A theory of gravity does two things.  It specifies a free, though generally
nonlinear, dynamics of the gravitational field and it specifies the coupling
between the gravitational field and matter.  The latter determines both how
matter generates and how it responds to gravity.

In the case of general relativity these conceptually distinct components of
gravitation physics are discernible in the standard form of the Einstein
field equation,
\begin{equation}
G_{\mu \nu} = 8 \pi T_{\mu \nu}.
\end{equation}
The structure of the Einstein tensor, on the left-hand side, specifies the
general relativistic dynamics of the gravitational field itself.  The form
of the stress-energy tensor, on the right-hand side, specifies the way in
which matter acts as a source of gravity.

These same aspects of general relativity and other Lagrangian-based
theories of gravity are even more clearly discernible in the structure
of the action principles from which their field equations follow,
\begin{equation}
\delta \! \int ({\cal L}_{G} + {\cal L}_{NG}) \  d^4x = 0. \label{2}
\end{equation}
The gravitational part of the Lagrangian density, ${\cal L}_{G}$, is
distinguished by the fact that it depends only on gravitational fields.
Its variational derivative with respect to these fields defines their
dynamics.  The variational derivative with respect to the metric of
general relativity's gravitational Lagrangian density,
$\sqrt{-g} R / 16 \pi$, is $\sqrt{-g} G_{\mu \nu}/16 \pi$.
The corresponding derivative of its nongravitational Lagrangian density
is $\sqrt{-g} T_{\mu \nu}/2$ and, so, the form of
${\cal L}_{NG}$, which depends on both matter and gravitational fields,
specifies the way in which matter generates gravity.  Its reciprocal role
of specifying the way in which matter responds to a gravitational field
follows from the action principle \ (\ref{2}) on variation with respect to
matter fields rather than gravitational ones.  This yields the matter
equations of motion.  In general, these need not follow directly from the
gravitational field equations as they do in general relativity.

Experimental tests of the Einstein equivalence principle focus on the behavior
of matter responding to a background gravitational field.  Their significance
lies in the way they constrain the form of ${\cal L}_{NG}$ and, thus, provide
an empirical foundation for one of gravitation physics' fundamental
components.  Tests that reveal no violation of the Einstein equivalence
principle force the form of the nongravitational Lagrangian density toward
ones that admit a metric representation, a representation in which a single
metric tensor field couples universally\cite{TLL} to all matter.

The experimental tests of the Einstein equivalence principle that we analyze
in this paper involve the propagation of polarized light through a background
gravitational field.  We are, therefore, concerned with only that part of
the nongravitational Lagrangian density which governs the dynamics of the
electromagnetic field and we base our analysis on a general model of this
Lagrangian density introduced by Ni\cite{Ni10}.  One advantage of this approach
is that the results of our analysis are easily specialized to yield the
predictions of any theory of gravity whose structure is encompassed by Ni's
$\chi g$ formalism.  The other advantage is that a broad consideration of
conceivable nonmetric couplings provides a context in which to judge how
different experimental tests of the Einstein equivalence principle compete
with and complement one another and, so, to clarify the strengths and
weaknesses of the empirical foundation of one of the most fundamental
aspects of gravitation physics.

The structure of the $\chi g$ formalism is consistent with the basic
tenets of the Dicke framework\cite{Dicke}. Spacetime is a differentiable
manifold and matter fields and phenomenological gravitational potentials
are tensor fields.  Given the apparent smoothness of space and time and
the view that physics should not depend in an essential way on a choice
of coordinates, it is difficult to fault these axioms as a basis for
phenomenology.  Furthermore, it is assumed that as gravity is ``turned off''
the nongravitational Lagrangian density, ${\cal L}_{NG}$, reduces smoothly
to the corresponding special relativistic Lagrangian density.  The most
general Lagrangian density consistent with these axioms and with assumptions
of electromagnetic gauge invariance, linearity of the electromagnetic field
equations and the absence of couplings to derivatives of gravitational
potentials is
\begin{equation}
{\cal L}_{EM} = - {1\over {16\pi}} \chi^{\alpha \beta \gamma \delta}
F_{\alpha \beta} F_{\gamma \delta},
\label{3}
\end{equation}
where the electromagnetic field is related to a 4-potential in the usual
way, $F_{\alpha \beta} = A_{\beta ,\alpha} - A_{\alpha ,\beta}$ and where
$\chi^{\alpha \beta \gamma \delta}$ is a tensor density that provides a
phenomenological representation of the gravitational field.  Note the
analogy between $\chi^{\alpha \beta \gamma \delta}$ and a macroscopic
electrodynamic constitutive tensor density.

To see how the phenomenological field $\chi^{\alpha \beta \gamma \delta}$ is
constructed from a specific theory's scalar, vector and tensor gravitational
fields, the structure of the theory's electromagnetic Lagrangian density can
be matched to that of ${\cal L}_{EM}$ in Eq.\ (\ref{3}).  In metric theories
of gravity it is constructed from the metric tensor alone,
$\chi^{\alpha \beta \gamma \delta} = {1\over 2} \sqrt{-g}
(g^{\alpha \gamma} g^{\beta \delta} - g^{\alpha \delta} g^{\beta \gamma})$.

Notice that $\chi^{\alpha \beta \gamma \delta}$ inherits symmetries analogous
to those of the Riemann tensor from the antisymmetry of $F_{\alpha \beta}$
and from the symmetrical way in which it appears in Eq.\ (\ref{3}).
Consequently, $\chi^{\alpha \beta \gamma \delta}$ has twenty-one
independent components.  Ni says that his formalism has twenty-one nonmetric
degrees of freedom because, in general, one cannot exploit the formalism's
covariance to transform away any of these independent component potentials.
The freedom to redefine coordinates is exhausted by putting the formalism's
$g_{\alpha \beta}$ field into a standard form.  This field represents the
coupling between the gravitational field and scalar particles.

\section{Geometric Optics in a Background Gravitational Field}

In this paper we consider the propagation of light through relatively
weak gravitational fields that vary on length and time scales that are
far longer than the light's wavelength and period respectively.  We,
therefore, adapt the coordinate system in which the Lagrangian
density (\ref{3}) is represented to the weak-field limit and we employ the
methodology of geometric optics to analyze the light's propagation.

In a Lorentzian coordinate system the electromagnetic Lagrangian density
of special relativity has the form in Eq.\ (\ref{3}) with
$\chi^{\alpha \beta \gamma \delta} \equiv {1\over 2}
(\eta^{\alpha \gamma} \eta^{\beta \delta}
- \eta^{\alpha \delta} \eta^{\beta \gamma})$.  Consequently, in the
weak field limit we can find quasi-Lorentzian coordinate systems in which
the field $\chi^{\alpha \beta \gamma \delta}$ has the form
\begin{equation}
\chi^{\alpha \beta \gamma \delta} \equiv {1\over 2}
(\eta^{\alpha \gamma} \eta^{\beta \delta}
- \eta^{\alpha \delta} \eta^{\beta \gamma})
+ \delta \! \chi^{\alpha \beta \gamma \delta},
\label{4}
\end{equation}
with $\delta \! \chi^{\alpha \beta \gamma \delta} \ll 1$.  Such a
decomposition of $\chi^{\alpha \beta \gamma \delta}$ is covariant
with respect to Poincar\'e transformations provided that boost velocities
too close to unity are not allowed.  Note that this is a purely formal
covariance.  The background gravitational field may well single out
a preferred frame\cite{Dicke}.

Geometric optics treats the propagation of locally plane electromagnetic
waves.  The amplitude and phase representation of such a wave,
\begin{equation}
{\bf E} = {\bf A}_E e^{i\Phi}, \qquad {\bf B} = {\bf A}_B e^{i\Phi}.
\label{5}
\end{equation}
is characterized by the fact that derivatives of the vector amplitudes
${\bf A}_E$ and ${\bf A}_B$ are small compared to derivatives of the
rapidly varying phase function $\Phi$.  Since we are interested in the
propagation of high-frequency electromagnetic radiation through a weak
background gravitational field that varies slowly in space and time,
derivatives of the background field are also small compared to derivatives
of $\Phi$.  We base the analysis that follows on the eikonal equation which
determines local coordinate velocities of wave propagation.  This equation
is derived by inserting the representation (\ref{5}) into the electromagnetic
field equations and ignoring all derivatives other than those of the phase
function.  Propagation equations for the vector amplitudes ${\bf A}_E$ and
${\bf A}_B$ and equations governing post-geometric-optic corrections
can be derived by taking systematic account of smaller derivatives but we
shall not need these equations here.

Neglecting derivatives of the slowly varying background field, the
electromagnetic field equation that follows from Eq.\ (\ref{3}) is
\begin{equation}
\chi^{\alpha \beta \gamma \delta} F_{\gamma \delta , \beta} = 0 .
\end{equation}
Defining electric and magnetic fields via $F_{i0} \equiv E_i$ and
$F_{jk} \equiv \epsilon_{jkl} B_l$ and employing the decomposition (\ref{4}) of
$\chi^{\alpha \beta \gamma \delta}$, this can be written somewhat more
transparently as
\begin{equation}
{\bf \nabla} \cdot {\bf E} +
{\rm terms\ proportional\ to\ } \delta \! \chi \ {\rm and\ } {\bf E}
\ {\rm or\ } {\bf B} = 0,
\label{7}
\end{equation}
and
\begin{equation}
{\bf \nabla} \times {\bf B} - { {\partial {\bf E}} \over {\partial t} } +
{\rm terms\ proportional\ to\ } \delta \! \chi \ {\rm and\ } {\bf E}
\ {\rm or\ } {\bf B} = 0.
\label{8}
\end{equation}
The electric and magnetic fields also satisfy
\begin{equation}
{\bf \nabla} \times {\bf E} + {{\partial {\bf B}}\over {\partial t}}=0,
\label{9}
\end{equation}
and
\begin{equation}
{\bf \nabla} \cdot {\bf B} = 0,
\label{10}
\end{equation}
by virtue of the fact that $F_{\alpha \beta}$ and, so, ${\bf E}$ and ${\bf B}$
are derived from potentials in the usual way.

To derive the eikonal equation we insert the representation (\ref{5}) of a
locally plane wave into the field equations (\ref{7}) - (\ref{10}) and
neglect all derivatives other than those of the rapidly varying phase
function $\Phi$.  Let $k_{\mu}$ denote the gradient of this function,
$k_{\mu} \equiv \partial_{\mu} \Phi
\equiv (\partial \Phi / \partial t, {\bf \nabla} \Phi)
\equiv (-\omega, {\bf k})$, so that Eqs.\ (\ref{9}) and (\ref{10}) become
\begin{equation}
{\bf A}_B = {{{\bf k} \times {\bf A}_E} \over \omega}
\label{11}
\end{equation}
and
\begin{equation}
{\bf k} \cdot {\bf A}_B = 0.
\end{equation}
The latter of these implies that the magnetic field of a locally plane wave
is transverse to the direction in which the wave propagates.  Since
Eq.\ (\ref{7}) becomes
\begin{equation}
{\bf k} \cdot {\bf A}_E =
{\rm terms\ proportional\ to\ } \delta \! \chi \ {\rm and\ }
{\bf A}_E \ {\rm or\ } {\bf A}_B,
\label{13}
\end{equation}
the electric field is guaranteed to be purely transverse only in the
absence of gravity.  To first order in the small
$\delta \! \chi^{\alpha \beta \gamma \delta}$ that represent the weak
background gravitational field, Eq.\ (\ref{13}) expresses a wave's longitudinal
electric field component in terms of its dominant transverse field
components.  To this same order, $O(\delta \! \chi )$, Eqs.\ (\ref{11})
and (\ref{13}) imply
\begin{equation}
{\bf A}_E = - {\omega \over k^2} {\bf k} \times {\bf A}_B +
{\rm terms\ proportional\ to\ } \delta \! \chi \ {\rm and\ } {\bf A}_B,
\label{14}
\end{equation}
which expresses ${\bf A}_E$ in terms of the two independent components of
${\bf A}_B$, and Eqs.\ (\ref{8}), (\ref{11}) and (\ref{14}) imply the
eikonal equation
\begin{equation}
(1 - {\omega^2 \over k^2}) {\bf A}_B =
{\rm terms\ proportional\ to\ } \delta \! \chi \ {\rm and\ } {\bf A}_B.
\label{15}
\end{equation}
Since the magnetic amplitude ${\bf A}_B$ has two independent components and
since $\omega / k$ is coordinate phase velocity, finding the two independent
polarization states that propagate with well-defined phase velocities is a
matter of solving a two-dimensional eigenvalue problem.  Note that in the
absence of gravity the right-hand side of Eq.\ (\ref{15}) vanishes implying,
as one expects, $\omega / k = 1$ regardless of polarization.

Before examining the explicit form of Eq.\ (\ref{15}) it is convenient to
make a 3+1 decomposition of $\delta \! \chi^{\alpha \beta \gamma \delta}$
into a set of SO(3) tensor objects.  We define
\begin{equation}
\xi^{ij} = -\delta \! \chi^{0i0j}, \qquad
\gamma^{ij} = {1\over 2} \epsilon^{jlm} \delta \! \chi^{0ilm},
\qquad {\rm and} \
\zeta^{ij} = {1\over 4} \epsilon^{ilm} \epsilon^{jpq} \delta \! \chi^{lmpq},
\end{equation}
where $\epsilon^{ijk}$ is the Levi-Civita antisymmetric symbol.  This
decomposition is covariant under rotations of the quasi-Lorentzian
coordinate system in which $\delta \! \chi^{\alpha \beta \gamma \delta}$
is represented. It follows that the tensors $\xi^{ij}$, $\gamma^{ij}$
and $\zeta^{ij}$ represent spatial anisotropy induced by the background
gravitational field in a natural way.  The symmetries of
$\chi^{\alpha \beta \gamma \delta}$
and $\delta \! \chi^{\alpha \beta \gamma \delta}$ imply that $\xi^{ij}$ and
$\zeta^{ij}$ are symmetric.  Their twelve independent components combine
with the nine of $\gamma^{ij}$ to account for the twenty-one nonmetric
degrees of freedom of the $\chi g$ formalism.  An analogous decomposition
of $\chi^{\alpha \beta \gamma \delta}$ yields tensors
$\epsilon^{ij} = \delta^{ij} + \xi^{ij}$,
$\chi^{ij} = \delta^{ij} + \zeta^{ij}$ and $\gamma^{ij}$.

Consider the propagation of light in the neighborhood of some event in
spacetime.  We exploit the covariance of our decomposition of
$\delta \! \chi^{\alpha \beta \gamma \delta}$ to rotate from the original
quasi-Lorentzian $(t,x,y,z)$ coordinate system in which the background
gravitational field is represented to a set of $(t,x',y',z')$ coordinates
in which the light propagates in the $z'$ direction.  In this new coordinate
system ${\bf A}_B$ has only $x'$ and $y'$ components so Eq.\ (\ref{15}) reduces
to the system of two equations,
\begin{equation}
(1 - {\omega^2 \over k^2}) A^{1'}_B = {\cal A} A^{1'}_B - {\cal B} A^{2'}_B,
\label{17}
\end{equation}
\begin{equation}
(1 - {\omega^2 \over k^2}) A^{2'}_B = -{\cal B} A^{1'}_B + {\cal C} A^{2'}_B.
\label{18}
\end{equation}
The coefficients ${\cal A}$, ${\cal B}$ and ${\cal C}$ depend on location in
spacetime and on the direction in which the wave propagates, the $z'$
direction.  This dependence is implicit in the expressions for
${\cal A}$, ${\cal B}$ and ${\cal C}$ in terms of the values at the event in
question of the tensor components $\xi^{i'j'}$, $\zeta^{i'j'}$ and
$\gamma^{i'j'}$ in the $(t, x',y',z')$ coordinate system,
\begin{equation}
{\cal A} = \xi^{2'2'} - 2 \gamma^{2'1'} - \zeta^{1'1'},
\end{equation}
\begin{equation}
{\cal B} = \xi^{1'2'} + (\gamma^{2'2'} - \gamma^{1'1'}) + \zeta^{1'2'},
\end{equation}
and
\begin{equation}
{\cal C} = \xi^{1'1'} + 2 \gamma^{1'2'} - \zeta^{2'2'} .
\end{equation}
The matrix defining the structure of the right-hand sides of Eqs.\ (\ref{17})
and (\ref{18}) is real-valued and symmetric when
$\delta \! \chi^{\alpha \beta \gamma \delta}$ is real.  Its eigenvalues are
\begin{equation}
\lambda_{\pm} = { {{\cal A} + {\cal C}} \over 2 }
\pm {1\over 2} \sqrt{({\cal A} - {\cal C})^2 + 4{\cal B}^2}.
\end{equation}
The corresponding eigenvectors define the polarization states that propagate
with well-defined coordinate phase velocities
$c_{\pm} = 1 - {1\over 2} \lambda_{\pm} + O(\delta \! \chi^2)$.

The effects of a nonmetric background field on physical standards of length
and time are such that a local observer who uses rulers and clocks to
measure velocities of light having the polarizations singled out by
the field will generally obtain results that differ from the coordinate
velocities $c_+$ and $c_-$ at $O(\delta \! \chi)$.  To that order, however,
the fractional difference between the velocities that an observer at rest
in the $(t,x',y',z')$ coordinate system does measure will be the same as the
fractional difference between $c_+$ and $c_-$.  We let $\delta c / c$ denote
the magnitude of this dimensionless local observable,
\begin{equation}
{{\delta c} \over c} = {1\over 2} \sqrt{({\cal A} - {\cal C})^2 + 4{\cal B}^2}
+ O(\delta \! \chi^2).
\label{23}
\end{equation}
A practical way to measure or to impose an upper limit on this magnitude is
to search for the effect such birefringence has on the polarization of light
that propagates through the background gravitational field.

We use Stokes parameters based on the linear polarization states singled
out by the background field to characterize the polarization of
light that propagates through a local observer's laboratory.  The
intensity $I$ and the degrees of polarization $Q/I$, $U/I$ and $V/I$
provide a particularly convenient representation of the light's state.
We may think of the light as an incoherent superposition of appropriate
intensities of unpolarized light, linearly polarized light with $Q/I = \pm 1$
and elliptically polarized light with $(U^2 + V^2)^{1/2} /I = \pm 1$.
Propagation through the observer's laboratory only affects the state of the
last of these components.

The unpolarized component is itself an incoherent superposition of equal
amplitudes of the basis polarization states while the linearly polarized
component consists purely of one or the other of the basis polarization
states.  The shift in the relative phase of waves in the basis
states that accumulates as these waves propagate with different phase
velocity has no effect on the polarization of either of these components.
On the other hand, the elliptically polarized component is a coherent
superposition of equal amplitudes of the basis states so its
polarization is sensitive to the relative phase of these waves.  Indeed,
the relative phase determines this component's $U/I$ and $V/I$ degrees of
polarization, $V/U = \tan (\Phi_+ - \Phi_- )$.  As light propagates
during a coordinate time interval $dt$ this relative phase shifts by
$\omega (\delta c / c) dt$, where $\omega$ is the light's angular frequency.
It is this shift that can affect light's polarization.

We piece together the cumulative effect of propagation through a background
gravitational field on a signal's polarization from such local effects.
This is particularly easy to do when the orientation of the basis
polarization states singled out by the background field is fixed along
light rays, as is the case for light propagating through the essentially
static, spherically symmetric background fields that we consider in the next
section.  A ray threading through a spherically symmetric field lies in a
plane and, for the fields we consider, there is a fixed orientation between
this plane and the ${\bf A}_B$ amplitudes of the polarization states singled
out by the fields.  In such cases the relative phase shift that accumulates
between waves in the basis states is simply
\begin{equation}
\Delta \Phi = \omega \int {{\delta c} \over c} dt
+ O(\delta \! \chi^2) ,
\label{24}
\end{equation}
where, since the local value of $\delta c / c$ is explicitly
$O(\delta \! \chi )$, the integration runs along the unperturbed
ray that runs straight through the quasi-Lorentzian coordinate system
with unit speed from source to receiver.  The integrand in Eq.\ (\ref{24})
is the local value of $\delta c / c$ for the ray's direction.

One consequence of the relative phase shift (\ref{24}) is that a signal whose
elliptically polarized component has $V/U = 0$ when emitted will have
some degree of circular polarization when it is received.  The degree of
circular polarization received is proportional to a small accumulated
shift,
\begin{equation}
V/U \approx \Delta \Phi .
\label{25}
\end{equation}
We base our interpretation of galaxy polarization data in the next section
on this expression.

We have seen that a nonmetric background field induces birefringence if and
only if $({\cal A} - {\cal C})$ or ${\cal B}$ in Eq.\ (\ref{23}) is
nonzero for some direction of wave propagation at some event in spacetime.
Before we turn to the derivation of empirical limits on such birefringence
let us consider how such limits constrain the form of the coupling between
gravitational and electromagnetic fields.

Note that the expressions $({\cal A} - {\cal C})$ and ${\cal B}$ involve only
the symmetric, trace-free parts of the tensors $\xi^{ij}$, $\zeta^{ij}$ and
$\gamma^{ij}$.  This is most easily appreciated by writing
$({\cal A} - {\cal C})$ and ${\cal B}$ in terms of the spherical components of
these tensors.  In terms of components in the $(t, x' ,y' ,z')$ coordinate
system they are
\begin{equation}
{\cal A} - {\cal C} = { 2 \over \sqrt{6} }
[ ( \xi^{(2)}_{2'} + \xi^{(2)}_{-2'} )
+ 2 i ( \gamma^{(2)}_{2'} - \gamma^{(2)}_{-2'} )
+ ( \zeta^{(2)}_{2'} + \zeta^{(2)}_{-2'} ) ]
\label{26}
\end{equation}
and
\begin{equation}
{\cal B} = - { 1 \over \sqrt{6} }
[ i ( \xi^{(2)}_{2'} - \xi^{(2)}_{-2'} )
+ 2 ( \gamma^{(2)}_{2'} + \gamma^{(2)}_{-2'} )
+ i ( \zeta^{(2)}_{2'} - \zeta^{(2)}_{-2'} ) ] .
\label{27}
\end{equation}
Only $l=2$ components appear.  Expressions for $({\cal A} - {\cal C})$ and
${\cal B}$ in terms of components in the original $(t,x,y,z)$ coordinate
system follow from the transformation law for spherical tensor
components\cite{Edmonds}, for example,
\begin{equation}
\xi^{(l)}_{m'} = {\cal D}^{(l)}_{m' m}(\phi, \theta, \psi) \xi^{(l)}_{m},
\label{28}
\end{equation}
where $\phi$, $\theta$ and $\psi$ are the Euler angles specifying the
rotation from $(t,x,y,z)$ to $(t,x',y',z')$.
 From the form of this transformation
law and of Eqs.\ (\ref{26}) and (\ref{27}) we conclude that the absence of
gravity-induced birefringence implies
$\xi^{(2)}_m + \zeta^{(2)}_m = 0 $ and $\gamma^{(2)}_m = 0$ and, so, that
observations imposing limits on the strength of gravity-induced
birefringence constrain ten of the $\chi g$ formalism's twenty-one nonmetric
degrees of freedom.  This conclusion and the expression (\ref{23}) for
$\delta c / c$ are in accord with remarks of Ni\cite{Ni8} for which the
preceding analysis provides a detailed justification.

\section{Gravity-Induced Birefringence and the Polarization of Galaxies}

The preceding analysis examined nonmetric couplings between the
gravitational and electromagnetic fields and established which
single out pairs of linear polarization states that propagate
through a gravitational field with different phase velocities.
Using Stokes parameters based on those states, we found that such
gravity-induced birefringence causes $U$-polarized radiation
to develop a degree of circular polarization as it propagates.

Here, we interpret polarization measurements of extragalactic sources
cataloged by Weiler and De Pater\cite{de Pater} in light of this prediction and
derive a precise new constraint on the strength of birefringence that
could be induced by the Galaxy's gravitational field.   Note that even a
cursory examination of the data indicates that such birefringence
cannot be strong.  If it were, one would expect to measure substantial
circular and linear polarization of radiation from extragalactic sources
with about the same likelihood.  This is not the case.   Measurements reveal
significant circular polarization of radiation from extragalactic sources
only rarely while they frequently reveal substantial linear polarization.

As a first approximation, we treat the Galaxy's gravitational field as
static and spherically symmetric.  This is not unreasonable, especially
if one accepts the existence of a substantial dark-matter halo.  Note
that the assumed spherical symmetry does not preclude the detection
of gravity-induced birefringence.  Radiation reaching the Earth from an
extragalactic source generally does not propagate radially through the
Galaxy's field because the Earth lies 10 kiloparsecs from the
Galaxy's center.  A ray along which such radiation travels lies in a
plane defined by the locations of the source, the Earth and the center
of the Galaxy.

At each point along a ray the only direction that the Galaxy's spherical
field can single out is the radial one.  We, therefore, find it useful to
introduce at each of these points a local quasi-Lorentzian $(t,x,y,z)$
coordinate system oriented so that its $z$ axis is radial and its $x$ axis
lies in the ray's plane.  We do so because the representation of the
spherical tensors introduced in the preceding section is simple in
these local systems.  Specifically, $\xi^{(2)}_m$, $\zeta^{(2)}_m$ and
$\gamma^{(2)}_m$ are nonzero only for $m=0$.  The nonzero components
depend only on the radial coordinate of the point along the ray.

At each of these points we can rotate the local $(t,x,y,z)$ coordinate
system about its $y$ axis through an angle $\theta$ to obtain a local
$(t,x',y',z')$ coordinate system in which the ray runs in the $z'$
direction.  Equation (\ref{23}) expresses the local value of $\delta c / c$
for points on the ray in terms of $({\cal A} - {\cal C})$ and ${\cal B}$
which are, in turn, expressed in terms of the local $(t,x',y',z')$
components $\xi^{(2)}_{\pm2'}$, $\zeta^{(2)}_{\pm2'}$ and
$\gamma^{(2)}_{\pm2'}$ by Eqs.\ (\ref{26}) and (\ref{27}).  Since the Euler
angles of the rotation from $(t,x,y,z)$ to $(t,x',y',z')$ are $\phi = 0$,
$\theta$ and $\psi = 0$, the transformation law (\ref{28}) implies
\begin{equation}
\xi^{(2)}_{\pm 2'} = \sin^2 (\theta ) \xi^{(2)}_0 ,
\label{29}
\end{equation}
with the same relationship holding between $\zeta^{(2)}_{\pm 2'}$ and
$\zeta^{(2)}_0$ and between $\gamma^{(2)}_{\pm 2'}$ and $\gamma^{(2)}_0$.
For any spherically symmetric gravitational field, we conclude that
$\delta c / c$ is proportional to the square of the sine of the angle
by which a ray's direction differs from the radial one.

The orientations of the ${\bf A}_B$ amplitudes of the polarization states
singled out along a ray are determined by the relative magnitudes of the
$(t,x,y,z)$ components $\xi^{(2)}_0$, $\zeta^{(2)}_0$ and $\gamma^{(2)}_0$.
The form of Eqs.\ (\ref{26}) and (\ref{27}) and the transformation law
(\ref{29}) imply that ${\cal B}$ is proportional to $\gamma^{(2)}_0$ while
$({\cal A} - {\cal C})$ is proportional to $(\xi^{(2)}_0 + \zeta^{(2)}_0)$.
In the event that $({\cal A} - {\cal C}) \neq 0$ and ${\cal B} = 0$,
Eqs.\ (\ref{17}) and (\ref{18}) imply that for the polarization states
singled out by the Galaxy's field
${\bf A}_B$ will lie in and perpendicular to the ray's plane.  On the other
hand, when $({\cal A} - {\cal C}) = 0$ and ${\cal B} \neq 0$, Eqs.\ (17) and
(18) imply the ${\bf A}_B$ singled out will be oriented at 45 degrees to
either side of the normal to the ray's plane.  Cases like these, in which
the polarization states singled out by a background field are fixed along
rays, are special.  Generally, their orientations will change along a ray
as the relative magnitudes of $\xi^{(2)}_0$, $\zeta^{(2)}_0$ and
$\gamma^{(2)}_0$ vary.

The effect that propagation through the Galaxy's gravitational field has
on the polarization of light from extragalactic sources clearly depends
on the global structure of that gravitational field.  We choose to
compute the effect for an NGT model of the Galaxy's field studied by
Krisher\cite{Krisher} so that we can compare the precision of the constraint
we impose on gravity-induced birefringence with the one he imposed.
This NGT field has $\xi^{(2)}_0 = 0$, $\zeta^{(2)}_0 = L^2(r)$ and
$\gamma^{(2)}_0 = 0$, where the function $L^2(r)$ specifies the
structure of the antisymmetric tensor part of the Galaxy's
nonsymmetric field.  The fact that only $\zeta^{(2)}_0$
is nonzero is a consequence of the nonmetric coupling between the
nonsymmetric field and the electromagnetic field having been tuned to
satisfy constraints imposed by other tests of the Einstein equivalence
principle\cite{Mann}.  Since only $\zeta^{(2)}_0$ is nonzero, the orientations
of the basis polarization states singled out by the Galactic field are
fixed along rays with their magnetic amplitudes lying in and perpendicular
to ray planes.  Consequently, Eq.\ (\ref{24}) gives the relative phase shift
that accumulates between waves in these polarization states as they
propagate.  We have $\delta c / c = {1 \over 2} L^2 \sin^2 (\theta )$, which
is consistent with the results of
Gabriel\cite{Gabriel1,Gabriel3} {\it et al}.\ since they denote
$L^2$ by $\Omega $.

In Krisher's model the antisymmetric part of the Galaxy's nonsymmetric
gravitational field is generated by a uniform density, spherical halo
of weakly interacting particles.  Its radius, $R$, is 25 kiloparsecs.
The function $L^2 (r)$ is proportional to the halo's net NGT charge, $l^2$.
We have $L^2 = l^4/r^4$ for radii greater than $R$ and $L^2 = l^4 r^2 / R^6$
for smaller radii.  The NGT charge, $l^2$, can be expressed in terms
of a coupling constant $f_c^2$ and the density of halo particles, $n_c$,
which Krisher takes to be 0.1 per cubic centimeter.  Specifically,
$l^2 = (4 \pi R^3 /3 ) n_c f_c^2 $.

The geometry of a ray that runs from an extragalactic source to the Earth
is shown in Fig.\ 1.  The spherical symmetry of the Galaxy's field
implies an axial symmetry of the relative phase shift (\ref{24}).  It depends
only on the angle $\beta$.  The 10 kiloparsec distance between the Earth
and the Galaxy's center is denoted by $d$.  The ray's impact parameter is
$b = d \sin(\beta)$.  To compute the phase shift (24), we parametrize the
rays so that the point $r = b$, which may lie on an extension of the ray,
corresponds to $t=0$ and we suppose that the ray starts from $r = \infty$ at
$t = - \infty$.  It follows that the ray encounters the halo's surface at
$t = t_R \equiv - (R^2 -b^2)^{1/2}$ and the Earth at
$t = t_E \equiv -d\cos(\beta)$.  Note that $r^2 = (b^2 + t^2)$ and
$\sin^2 (\theta) = b^2 / (b^2 + t^2)$ along the ray.

 From these facts and the expressions for $L^2$ inside and outside the halo
we conclude that
\begin{equation}
\Delta \Phi_{ext} = { {\omega l^4} \over 2}
\int_{-\infty}^{-t_R} { {b^2 dt} \over (b^2 +t^2)^3 }
= { {\omega l^4} \over 8 } \bigl( {t \over {(b^2 + t^2)^2}}
+ {{3 t} \over { 2 b^2 (b^2 + t^2)}}
+ {3 \over {2b^3}} \arctan ({t\over b})\bigr) \big|^{t_R}_{-\infty}
\label{30}
\end{equation}
and
\begin{equation}
\Delta \Phi_{int} = { {\omega l^4 b^2} \over {2 R^6} }
\int_{t_R}^{t_E} dt
= { {\omega l^4 b^2} \over {2 R^6} } ( t_E - t_R ).
\end{equation}
Equation (\ref{30}) corrects a misprint in Krisher's expression for
$\Delta \Phi_{ext}$.
More seriously, Krisher's numerical estimate of $\Delta \Phi_{ext}$ indicates
that $t_E$ rather than $t_R$ was used as an upper limit of integration.  This
contribution to $\Delta \Phi$ was, therefore, significantly overestimated.
In Fig.\ 2 we plot conveniently normalized functions $F_{ext} (\beta)$ and
$F_{int} (\beta)$ defined by $\Delta \Phi_i = (\omega l^4 / d^3 ) F_i$.

Data from the catalog of Weiler and De Pater impose a sharp new limit on
the amplitude of $\Delta \Phi (\beta)$ and, so, on the magnitudes of
$l^2$ and $f_c^2$.

 From the position of a source in the sky and a bit of geometry we obtain
both the angle $\beta$ for the source and the orientation of the
polarization states singled out for it by the Galaxy's field.  We use
this latter information to express the source's cataloged polarization
in terms of Stokes parameters based on the polarization states singled
out.  Since the relative phase shift $\Delta \Phi$ converts $U$ polarized
light to $V$ polarized light, we calculate $V/U$ for sources with
accurately measured $U/I$ and $V/I$ and for which the $U/I$ degree of
polarization is substantial, accounting for more than half of the
observed linear polarization.  The substantial $U/I$ degree of
polarization means that a given relative phase shift will induce a
relatively large $V/I$ and it also assures that $V/U$ will be a
well-behaved indicator of $\Delta \Phi$ via Eq.\ (\ref{25}) where we make
the conservative assumption that a nonzero average of $V/U$ for
sources in an interval around $\beta$ represents the effect of the
phase shift $\Delta \Phi (\beta)$.  Figure 3 is a plot of the $V/U$
as a function of $\beta$ for data taken at 5 GHz, the highest frequency
included in the Weiler-De Pater catalog.  The average of $V/U$ for
these sources is consistent with zero.  We estimate limits imposed
by this data by noting that 0.12 is the sample standard
deviation of $V/U$ for the 5GHz data having $\beta > 60$ degrees.
Equation (\ref{25}) imposes this as a limit on the mean of $\Delta \Phi$
for $\beta$ in this range.  This limit and the expression in terms of
$f_c^2$ for the mean of $\Delta \Phi (\beta)$ in the range from 60 to
90 degrees implies $f_c^2 < 1.4 \times 10^{-34}$ cm$^2$.  When the
means and sample standard deviations of $V/U$ for lower frequency data
are accounted for this constraint is sharpened slightly,
\begin{equation}
f_c^2 < 1.3 \times 10^{-34} cm^2.
\label{32}
\end{equation}

This limit is more than $10^4$ times tighter than the one Krisher derives,
once his overestimate of the exterior contribution to $\Delta \Phi$ is
corrected for.  This increase in precision is easily understood.  Krisher
argued that NGT-induced birefringence would affect the polarization structure
of pulses from pulsars in the Large and Small Magellenic Clouds and reckoned
that an effect would have been noticed if the arrival times of pulse
components in the polarization states singled out by the Galaxy's field
were shifted relative to each other by as much as a millisecond.  In our
case, we use a kind of interferometry to discern a much smaller relative
time delay.  The highest frequency data that we analyze imposes a limit
of 0.12 radians on the relative phase shift between the polarization states
singled out by the Galaxy's field.  We are, therefore, resolving a fraction
$0.12 / 2 \pi$ of a cycle at the 5 GHz observation frequency.  This
corresponds to a relative time delay of $4 \times 10^{-12}$ seconds.  Our
time resolution is, therefore, more than $10^8$ times Krisher's.  Since,
as Krisher remarks, constraints on the magnitudes of $l^2$ or $f_c^2$ are
proportional to the square root of the time interval resolved, the
precision of our constraint (\ref{32}) is accounted for.

\section{Summary and Conclusions}

The local isotropy of space is one consequence of the Einstein equivalence
principle.  Atomic physics experiments of Hughes-Drever\cite{Hughes-Drever}
type are the classic tests of this prediction.  In this paper we have shown
that searching for effects of propagation through a gravitational field on
the polarization of light provides a precise, qualitatively different test
for spatial anisotropy, specifically spatial anisotropy that induces
a birefringence of space.  In the context of the $\chi g$ formalism such
anisotropy is associated with ten of twenty-one nonmetric degrees of
freedom.

The significance of any test of the Einstein equivalence principle lies in
the constraint it imposes on the form of the coupling between gravitational
and matter fields.  Constraints on gravity-induced birefringence force the
structure of the electromagnetic Lagrangian density (\ref{3}) towards the
form having
\[
\chi^{\alpha \beta \gamma \delta} = {1\over 2} \sqrt{-\tilde g}
( \tilde g^{\alpha \gamma} \tilde g^{\beta \delta}
- \tilde g^{\alpha \delta} \tilde g^{\beta \gamma} )
+ \phi \epsilon^{\alpha \beta \gamma \delta}.
\]
That the phenomenological symmetric tensor potential $\tilde g^{\alpha \beta}$
and scalar potential $\phi$ can represent the $\chi g$ formalism's
remaining eleven nonmetric degrees of freedom is clear upon counting
components.  The fact that this Lagrangian density predicts no birefringence
in the geometric optics limit is clear from the metric form of the coupling
between $\tilde g^{\alpha \beta}$ and the electromagnetic field and from the
identification of $\phi$ as the $\gamma^{(0)}_0$ part of
$\delta \! \chi^{\alpha \beta \gamma \delta}$. Note, however, that
Carroll and Field\cite{Carroll} have shown that at post-geometric-optics
order the $\phi$ coupling does induce a dispersive optical activity which
they have constrained using galaxy polarization data.

Two features are primarily responsible for the precision of Einstein
equivalence principle tests based on the interpretation of polarization
data for extragalactic sources.  First, they focus on global consequences
of Einstein equivalence principle violations.  They constrain shifts in
relative phase that accumulate from minute local effects rather than seeking
the local effects directly.  Second, they are interferometric in character.
Their sensitivity to accumulated phase shifts comes from exploiting the
coherence inherent in polarized light.  Note that the sharp new constraint
on NGT derived above is far beyond the reach of local Hughes-Drever
experiments, as indicated by an estimate analogous to one that
Gabriel\cite{Gabriel1} {\it et al}.\ make of local anisotropy induced by the
Sun's field in NGT.

Sharper tests of the Einstein equivalence principle are all but certain
to result from surveys of extragalactic sources that yield linear and circular
polarization measurements at frequencies higher than the 5 GHz limit of the
Weiler-De Pater catalog.

\begin{figure}
\caption{Points $G$, $E$, and $S$ depict the locations of the Galaxy's center,
the Earth and an extragalactic source. A halo of radius $R = 25 kpc$
surrounds the galaxy. The Earth lies at a distance $d = 10 kpc$ from $G$.
Light from the source $S$  propagates to the Earth along the ray $SE$.
The relative phase shift that accumulates depends only on the angle $\beta$
or, equivalently, on the impact parameter $b = d\sin(\beta)$. At each point
$P$ of $SE$ we denote the angle between $SE$ and the radial line $GP$
by $\theta(P)$.}
\label{fig1}
\end{figure}
\begin{figure}
\caption{The function of $\beta$ depicted by the solid curve is proportional
to the relative phase delay. It is the sum of the internal and external
contributions, $F_{int}(\beta)$ and $F_{ext}(\beta)$, defined in the text and
denoted here by dotted and dashed curves respectively.}
\label{fig2}
\end{figure}
\begin{figure}
\caption{Plot of $V/U$ ${\it versus}$ $\beta$ for  5GHz data from the
Weiler-De Pater catalog [11]. The apparent clustering of data is a consequence
of including data taken at different times for some sources.}
\label{fig3}
\end{figure}

\begin{thebibliography}{99}
\bibitem{Gabriel1}  M.\ D.\ Gabriel, M.\ P.\ Haugan, R.\ B.\ Mann and
J.\ H.\ Palmer, Phys.\ Rev.\ D{\bf 43}, 308 (1991) and
Phys.\ Rev.\ D{\bf 43}, 2465 (1991).
\bibitem{Moffatt}  J.\ W.\ Moffat in {\it Gravitation: A Banff Summer
Institute} edited by R.\ B.\ Mann and P.\ Wesson (World Scientific,
Singapore,1991) and references therein.
\bibitem{Gabriel3}  M.\ D.\ Gabriel, M.\ P.\ Haugan, R.\ B.\ Mann and
J.\ H.\ Palmer, Phys.\ Rev.\ Lett.\ {\bf 67}, 2123 (1991).
\bibitem{Krisher}  T.\ P.\ Krisher, Phys.\ Rev.\ D{\bf 44}, R2211 (1991).
\bibitem{Hughes-Drever}  V.\ W.\ Hughes, H.\ G.\ Robinson and
V. Beltran-Lopez, Phys.\ Rev.\ Lett.\ {\bf 4}, 342 (1960);
R.\ W.\ P.\ Drever, Phil.\ Mag.\ {\bf 6}, 683 (1960);
J.\ D.\ Prestage {\it et al}., Phys.\ Rev.\ Lett.\ {\bf 54}, 2387 (1985);
S.\ K.\ Lamoreaux {\it et al}., Phys.\ Rev.\ Lett.\ {\bf 57}, 3125 (1986)
and T.\ E.\ Chupp {\it et al}., Phys.\ Rev.\ Lett.\ {\bf 63}, 1541 (1989).
\bibitem{Dicke} C.\ M.\ Will, {\it Theory and Experiment in Gravitation
Physics}, (Cambridge University Press, Cambridge, England, 1981).
\bibitem{Ni8}  W.\ - T.\ Ni in {\it Precision Measurements and Fundamental
Constants II}, editied by B.\ N.\ Taylor and W.\ D.\ Phillips, U.\ S.\
National Bureau of Standards Special Publication No.\ 617 (U.\ S.\ GPO,
Washington, D.\ C.\ , 1984).
\bibitem{TLL}  K.\ S.\ Thorne, D.\ L.\ Lee and A.\ P.\ Lightman,
Phys.\ Rev.\ D{\bf 7}, 3563 (1973).
\bibitem{Ni10} W.\ - T.\ Ni, Phys.\ Rev.\ Lett.\ {\bf 38}, 301 (1977).
\bibitem{Edmonds} A.\ R.\ Edmonds, {\it Angular Momentum in Quantum Mechanics},
(Princeton University Press, Princeton 1974).
\bibitem{de Pater} K.\ W.\ Weiler and I.\ De Pater,
Ap.\ J.\ Suppl.\ {\bf 52}, 293 (1983).
\bibitem{Mann} R.\ B.\ Mann, J.\ H.\ Palmer and J.\ W.\ Moffat,
Phys.\ Rev.\ Lett.\ {\bf 62}, 2765 (1989).
\bibitem{Carroll} S.\ M.\ Carroll and G.\ B.\ Field,
Phys.\ Rev.\ D{\bf 43}, 3789 (1991).
\end{thebibliography}
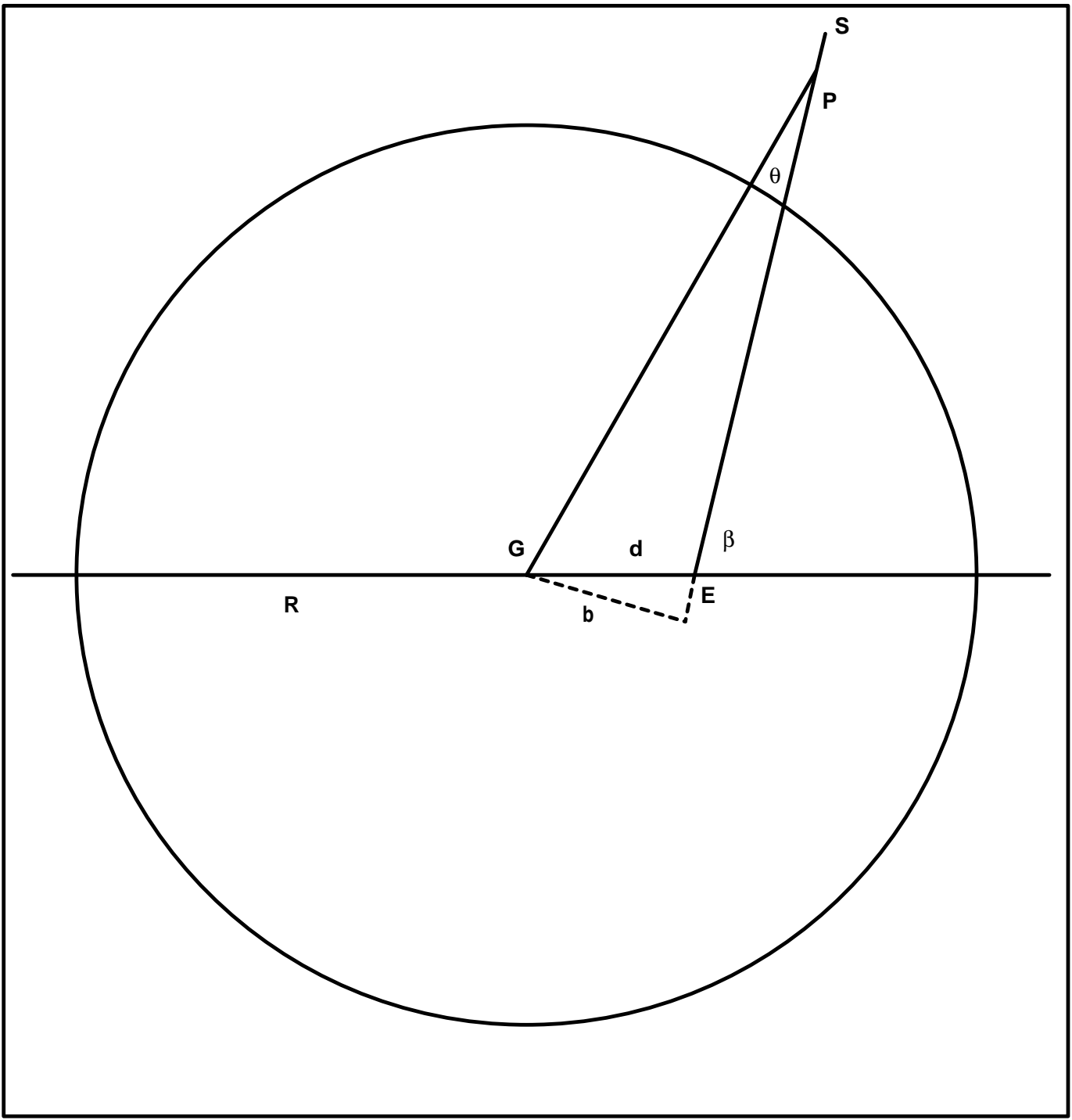
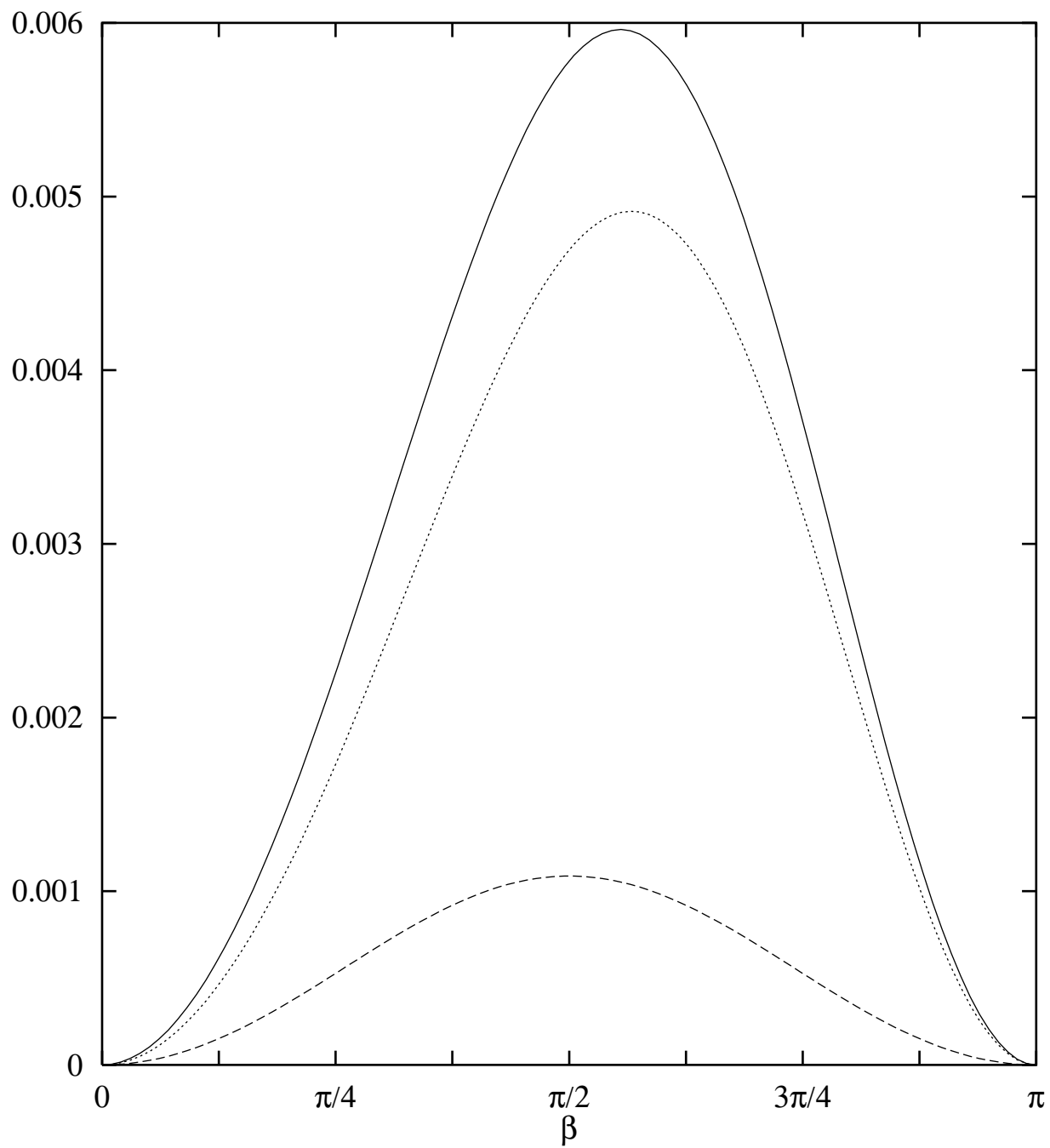
\end{document}